\documentclass[12pt]{article}
\usepackage{amsmath,amsfonts,amssymb,epsfig}
\usepackage{latexsym}
\def\Tr{{\rm{Tr}}}
\def\beq{\begin{equation}}
\def\eeq{\end{equation}}
\def\bea{{\begin{eqnarray}}}
\def\eea{\end{eqnarray}}
\def\bea*{\begin{eqnarray*}}
\def\eea*{\end{eqnarray*}}

\begin{document}

\begin{titlepage}

\begin{centering}

\vspace*{3cm}

\textbf{\Large Effects of a fundamental mass term in two-dimensional 
super Yang--Mills theory}

\vspace{1.5cm}

\textbf{Uwe Trittmann}

\vspace{0.5cm}

\textsl{Department of Physics \& Astronomy\\
Otterbein College\\
Westerville, OH 43081}
\vspace{1.0cm}

\textbf{Stephen S.~Pinsky}

\vspace{0.5cm}

\textsl{Department of Physics\\
The Ohio State University\\
Columbus, OH 43210}

\vspace{0.5cm}
%
%

\vspace{0.5cm}

\begin{abstract}
We show that adding a vacuum expectation value to a 
gauge field left over from a dimensional reduction of three-dimensional pure 
supersymmetric Yang-Mills theory generates mass terms for 
the fundamental fields in the two-dimensional theory while 
supersymmetry stays intact. This is similar to the adjoint mass term that is 
generated by a Chern-Simons term in this theory. We study the spectrum 
of the two-dimensional theory as a function of the 
vacuum expectation value
and of the Chern-Simons coupling.
Apart from some symmetry issues a straightforward picture arises. 
We show that at least one massless state exists if the Chern-Simons coupling 
vanishes.  
The numerical 
spectrum separates into (almost) massless and very heavy states as the  
Chern-Simons coupling grows. We present evidence that 
the gap survives the continuum limit. 
We display structure functions and other properties of some 
of the bound states. 
\end{abstract}

\end{centering}


\vfill

\end{titlepage}
\newpage
%
\section{Introduction}
\label{sectintro}

With the LHC experiments at CERN to begin data taking shortly, it may 
soon be clear 
whether supersymmetry is realized in Nature. Notwithstanding the experimental
verdict, supersymmetry provides solutions to profound questions in particle
physics \cite{SUSY2,SUSY}, and, as a symmetry, it is useful to simplify 
calculations. We use it within the framework of supersymmetric
discretized light-cone quantization (SDLCQ) 
to solve quantum field theories. SDLCQ comes with a set of strengths
and weaknesses documented in the literature \cite{DLCQ,SDLCQ}. In particular,
SDLCQ is a Hamiltonian approach and practically limited to theories with
enough supersymmetry to render them finite. It is primarily a numerical 
approach, and as such it is
cheaper to consider theories in lower dimensions. 
The SDLCQ Hamiltonian
is manifestly 
invariant under supersymmetry which is hard to achieve in conventional
lattice gauge theory because of the asymmetric treatment of bosons and 
fermions, although progress is being made \cite{Kaplan}.

In a line of publications, we have been deciphering 
the properties of bound states of theories that share
features with QCD or are interesting in their own right. Starting from 
a two-dimensional pure super Yang-Mills (SYM) theory 
\cite{Sakai,ImprovedSakai}, we have 
been adding fundamental matter to emulate quarks 
\cite{Lunin,FundMatter2D}, 
a Chern-Simons (CS) term to 
simulate effective gluon masses \cite{CS2D}, 
and tackled higher dimensional theories \cite{Haney,CS3D}. 
As a natural extension to previous work, we set out to construct 
a mass term for the fundamental particles in the present note.
Of course, supersymmetry itself prevents the use of ordinary mass terms, but
one does not have to think too hard to fix this problem.
Inspired by work of Myers, et al \cite{Myers} on ${\cal N}=2$ SYM in four 
dimensions, we might try to add a vacuum expectation value (VEV) to a 
gauge field left over from the dimensional reduction of a higher-dimensional 
version of the theory. Shifting the field by its VEV 
should then produce fundamental mass terms invariant under supersymmetry.
It turns out that the simplest scenario suffices: we can start with a 
three-dimensional ${\cal N}=1$ SYM theory, reduce it to two dimensions,
have the transverse gauge field acquire a VEV, 
and produce the desired fundamental mass term in the dimensionally reduced theory.   

In the present note, we work out the details and study the ensuing spectrum 
of bound states as a function both of the VEV (``quark mass'') and 
the CS coupling (``gluon mass''). 
Though the focus is on the effects of the VEV-induced mass terms, it is natural
to include a CS term, too.
We find that the theory containing both terms is not invariant
under any of the customary discrete symmetries. However, mass differences 
between nominal parity partners
are tiny due to a small symmetry-breaking term. Apart from this
glitch, exploring the model is straightforward and yields few surprises.
This is good news since stable continuum results can be extracted and a theory
with an interesting mass spectrum emerges. If no CS term is present 
massless states exist, otherwise the lightest states remain massive.
Masses tend to decrease with growing resolution, but even at finite $K$ some 
states become massless at special values of the VEV, if the CS coupling 
vanishes. As the couplings grow, few nearly massless states are
clearly separated from the bulk of heavy states. While it is hard to show 
directly that this feature survives the continuum limit, it is likely to be 
true at substantial CS coupling judging from the strong coupling limit 
and numerical evidence.
We present the theory in the next section, derive some analytical 
results, display numerical results, and conclude.

\section{Super Yang-Mills theory in two dimensions}
\label{sectSYM}
A two-dimensional super Yang-Mills theory with a Chern-Simons term is
generated conveniently by dimensionally reducing its three-dimensional pendant.
The action of $\mathcal{N}=1$ supersymmetric gauge theory in
three dimensions coupled to fundamental matter with a Chern--Simons
term is
\begin{equation}\label{Action}
S_{2+1}=S_{\rm SYM}+S_{\mathrm{fund}}+S_{\rm CS},
\end{equation}
with
\begin{eqnarray}
\label{eqn:SYM} S_{\rm SYM}&=&\int d^3x \,\Tr
\left(-\frac{1}{4}F_{\mu\nu}F^{\mu\nu} +
\frac{i}{2}{\bar\Lambda} \Gamma^\mu D_\mu \Lambda \right),\\
S_{\mathrm{fund}}&=&\int d^3x \, \left(D_\mu \xi^\dagger
D^\mu \xi+ i\bar{\Psi} D_\mu\Gamma^\mu\Psi -g\left[\bar{\Psi}\Lambda\xi+
\xi^\dagger{\bar\Lambda}\Psi\right]\right),\\
\label{eqn:SCS} S_{\rm CS}&=&\int d^3x \,
\frac{\hat{\kappa}}{2}\left(\epsilon^{\mu\nu\lambda}\left(
A_{\mu}\partial_{\nu}A_{\lambda}+ \frac{2 i}{3}g
A_{\mu}A_{\nu}A_{\lambda} \right)+ 2 \bar{\Lambda}\Lambda\right).
\end{eqnarray}
The gauge part, $S_{\rm SYM}$, of the action describes a system of gauge 
bosons $(A_\mu)_{ab}$ and
their superpartners, the Majorana fermions $\Lambda_{ab}$ with color indices
$a,b=1,\ldots, N_c$, transforming under the adjoint 
representation of $SU(N_c)$.  The matter content of the theory 
consists of a complex scalar $\xi_a$, and a Dirac fermion $\Psi_a$, both 
transforming under the fundamental representation of the 
gauge group. 
In matrix notation the
covariant derivatives and the gauge field strength are defined as usual
\begin{align}
\label{eqn:covdev}
D_\mu\Lambda&=\partial_\mu\Lambda+ig[A_\mu,\Lambda], &
D_\mu\xi&=\partial_\mu\xi+ig A_\mu\xi, &
D_\mu\Psi&=\partial_\mu\Psi+ig A_\mu\Psi, \\
D_\mu \xi^{\dag}&=\partial_\mu\xi^{\dag}-ig\xi^{\dag}A_\mu, &
D_\mu\Psi^{\dag}&=\partial_\mu\Psi^{\dag}-ig\Psi^{\dag}A_\mu, & 
F_{\mu\nu}&=\partial_{[\mu}A_{\nu]}+i g [A_{\mu}, A_{\nu}]. \nonumber
\end{align}
%
The action
(\ref{Action}) is invariant under supersymmetry transformations 
parameterized by a constant two-component
Majorana spinor $\varepsilon\equiv (\varepsilon_{1},\varepsilon_2)^{\rm T};$  
$\bar{\varepsilon}\equiv
\varepsilon^{\mathrm{T}}\Gamma^{0}$:
\begin{eqnarray}
\label{eqn:susyvars}
\delta A_\mu=\frac{i}{2} {\bar\varepsilon}\Gamma_\mu\Lambda,&&\quad
\delta\xi=\frac{i}{2}{\bar\varepsilon}\Psi, \quad
\delta\xi^{\dag}=-\frac{i}{2}\bar{\Psi}\varepsilon,\\
\delta\Lambda=\frac{1}{4}F_{\mu\nu}\Gamma^{\mu\nu}\varepsilon,&&\quad
\delta\Psi=-\frac{1}{2}\Gamma^\mu\varepsilon D_\mu\xi, \quad
\delta\bar{\Psi}=-\frac{1}{2}D_\mu\xi^{\dag}\bar{\varepsilon}\Gamma^\mu, \nonumber
\end{eqnarray}
where $\Gamma^{\mu\nu}=\frac{1}{2}\left[\Gamma^{\mu},\Gamma^{\nu}\right]$.
Using standard Noether techniques, we can determine the conserved current
density ${\cal J}^{\mu}=N^{\mu}+K^{\mu}$, consisting of the familiar Noether 
(on shell) current density $N^{\mu}$, and $K^{\mu}$, related to the 
change of the Lagrangian under a supersymmetry transformation and having 
the form of a space-time divergence. 
We dimensionally reduce the theory to two dimensions by omitting 
all transverse derivatives, 
$\partial_{\perp}(\ldots)=\partial^{\perp}(\ldots)\equiv 0$. Note that 
$A^{\perp}$ will remain part of the two-dimensional theory.

At this point it is useful to transcribe to 
light-cone coordinates, $x^\pm=(x^0 \pm x^1)/\sqrt{2}$. We calculate
the supercharge $Q^{\alpha}$ (a two-component spinor in two dimensions) 
by integrating the plus-component of the supercurrent ${\cal J}^+$ 
over all space, {\em i.e.}~over $x^-$ and $x^{\perp}$. The latter yields a constant 
factor which can be reabsorbed in a rescaling of the fields.
Using light-cone coordinates allows us to express the 
supercharge in terms of the {\em physical}
fields by imposing the light-cone gauge condition $A^+=A_-=0$. 
Since the other component of the gauge field can be eliminated by a constraint
equation, we are left with the adjoint $A^\perp_{ab}$ and the fundamental scalar 
$\xi_a$ as physical bosonic fields, whereas the (left-moving) physical fermionic 
fields $\lambda_{ab}$ and $\psi_a$ are components of the spinors appearing in
the action (\ref{Action})
\begin{equation}
\Lambda=\left(\begin{array}{c} \lambda\\ {\tilde\lambda}\end{array}\right),
\qquad
\Psi=\left({\psi \atop {\tilde\psi}}\right).
\end{equation}
We have used the imaginary (Majorana) representation
\begin{equation}
\Gamma^0=\sigma_2,\qquad \Gamma^1=i\sigma_1,\qquad \Gamma^{\perp}=i\sigma_3,
\end{equation}
to render the Majorana spinor field real, $\Lambda^{\dag}=\Lambda^{T}$. 
The supercharge components are labeled $Q=\left({Q^+, Q^-}\right)^T$, to 
reflect their relation to the Lorentz generators via the superalgebra in 
its $\mathcal{N}=(1,1)$ form
\begin{equation}
\label{eqn:sualg}
\{Q^+,Q^+\}=2\sqrt{2}P^+,\qquad \{Q^-,Q^-\}=2\sqrt{2}P^-,\qquad
\{Q^+,Q^-\}=0.
\end{equation}
%
%
The two-dimensional supercharge reads 
\beq\label{Supercharge}
Q^-=Q^-_{\rm SYM}+Q^-_{\rm fund}+Q^-_{\rm CS},
\eeq
where
\beq\label{QsymFields}
Q^-_{\rm SYM}={ig}\sqrt{2}\int dx^- \Tr\left[\left(-\left[A^{\perp},
\partial_-A^{\perp}\right]+\frac{i}{\sqrt{2}}\left\{\lambda,\lambda\right\}
\right)\frac{1}{\partial_-}\lambda\right],\nonumber
\eeq
\beq\label{QfundFields}
Q^-_{\rm fund}=ig\sqrt{2}\int dx^- \left(\Tr\left[\left(\partial_-\xi 
\xi^{\dagger}- \xi\partial_-\xi^{\dagger}+\sqrt{2} i\psi\psi^{\dagger}\right)
\frac{1}{\partial_-}\lambda\right]+i \xi^{\dagger}A^{\perp}\psi
+i \psi^{\dagger}A^{\perp}\xi\right),\nonumber
\eeq
\beq\label{QcsFields}
Q^-_{\rm CS}=\hat{\kappa}\sqrt{2}
\int dx^- \Tr\left(\lambda A^{\perp}\right). \nonumber
\eeq

To generate a mass term, we will assume\footnote{Introducing the VEV 
earlier would have required a modification of the super 
transformation of the fundamental fermion, entailing $\Psi/A$ mixing. 
It must, however, lead to the same supercharge and Hamiltonian.} 
that the gauge field $A^{\perp}_{ab}$ acquires 
a vacuum expectation value 
\[
\hat{v}_{ab}:=\langle A^{\perp}_{ab}\rangle = \hat{v} \delta_{ab}.
\]
Shifting the field by its VEV, and expressing the theory in terms of the 
new field 
\beq\label{shift} 
(A^{\perp}_{ab})\,'=A^{\perp}_{ab}-\langle A^{\perp}_{ab}\rangle,
\eeq
will yield extra terms in the supercharge, which can be interpreted as 
mass terms for the fundamental particles of the theory. 
The only part of the supercharge
that is affected by the shift of the perpendicular gauge field 
is $Q^-_{\rm fund}$, since the color-neutral, constant VEV 
appears in a derivative in $Q^-_{\rm SYM}$, and in a trace in $Q^-_{\rm CS}$.
The effect of the shift, Eq.~(\ref{shift}), is on the last two terms of 
$Q^-_{\rm fund}$, 
giving rise to an extra operator in the supercharge
\beq\label{QxsFields}
Q^-_{\rm XS}=-g\hat{v}\sqrt{2}\int dx^-
\left(\xi^{\dagger}\psi+\psi^{\dagger}\xi \right).
\eeq
At this point, we employ the framework of SDLCQ (see, e.g.~\cite{TDFundMatter}) 
to obtain the mode decomposition
of the supercharge listed in the Appendix that will allow us
to evaluate the theory on a computer. In particular,
we quantize by imposing the canonical commutation relations
\begin{eqnarray}
\left[A_{ab}^{\perp}(0,x^-),\partial_-A_{cd}^{\perp}(0,y^-)\right]&=&
i\delta_{ad}\delta_{bc}
\delta(x^--y^-)\,,\nonumber\\
\left\{\lambda_{ab}(0,x^-),\lambda_{cd}(0,y^-)\right\}&=&\sqrt{2}
\delta_{ad}\delta_{bc}\delta(x^--y^-)\,,\nonumber\\
\left[\xi_a(0,x^-),\partial_-\xi_b(0,y^-)\right]&=&
i\delta_{ab}\delta(x^--y^-)\,,\nonumber\\
\left\{\psi_{a}(0,x^-),\psi^{\dagger}_{b}(0,y^-)\right\}&=&\sqrt{2}
\delta_{ab}\delta(x^--y^-)\,.
\label{CanComRelField}
\end{eqnarray}
The compactification of the theory on a on a 
lightlike circle ($-L<x^-<L$) 
leads to discrete momentum modes defined via
\begin{eqnarray}
\label{expandA2}
A^{\perp}_{ab}(0,x^-)&=&\frac{1}{\sqrt{4\pi}}\sum_{n=1}^{\infty}\frac{1}{\sqrt{n}}
\left(A_{ab}(n)e^{-in\pi x^-/L}+A^\dagger_{ba}(n)e^{in\pi x^-/L}\right)\,,\\
\label{expandLambda}
\lambda_{ab}(0,x^-)&=&\frac{1}{2^{\frac{1}{4}}\sqrt{2L}}\sum_{n=1}^{\infty}
\left(B_{ab}(n)e^{-in\pi x^-/L}+B^\dagger_{ba}(n)e^{in\pi x^-/L}\right)\,,\\
\label{expandxi}
\xi_a(0,x^-)&=&\frac{1}{\sqrt{4\pi}}\sum_{n=1}^{\infty}\frac{1}{\sqrt{n}}
\left(C_a(n)e^{-in\pi x^-/L}+{\tilde C}^\dagger_{a}(n)e^{in\pi
x^-/L}\right)\,,\\
\label{expandPsi}
\psi_{a}(0,x^-)&=&\frac{1}{2^{\frac{1}{4}}\sqrt{2L}}\sum_{n=1}^{\infty}
\left(D_{a}(n)e^{-in\pi x^-/L}+{\tilde D}^\dagger_{a}(n)e^{in\pi
x^-/L}\right)\,.
\end{eqnarray}
Normalization is chosen such that the commutation relations 
(\ref{CanComRelField}) in terms of 
creation and annihilation operators of are cast into their customary
form, namely 
\beq
\left[A_{ab}(n),A^\dagger_{cd}(n')\right]=
\left\{B_{ab}(n),B^\dagger_{cd}(n')\right\}=
\left(\delta_{ad}\delta_{bc}-\frac{1}{N}\delta_{ab}\delta_{cd}\right)
\delta_{nn'}\,,
\eeq
\beq
\left[C_{a}(n),C^\dagger_{b}(n')\right]=
\left[{\tilde C}_{a}(n),{\tilde C}^\dagger_{b}(n')\right]=
\left\{D_{a}(n),{D}^\dagger_{b}(n')\right\}=
\left\{{\tilde D}_{a}(n),{\tilde D}^\dagger_{b}(n')\right\}
=\delta_{ab}\delta_{nn'}\,.\nonumber
\eeq
The extra part of the supercharge becomes 
\[
Q^-_{\rm XS}=-\frac{g\hat{v}}{2^{1/4}}\sqrt{\frac{L}{\pi}}
\sum_{n=1}^{\infty}\frac{1}{\sqrt{n}}
\left(C_a^{\dagger}(n)D_a(n)+\tilde{C}_a^{\dagger}(n)\tilde{D}_a(n)+
D_a^{\dagger}(n)C_a(n)+\tilde{D}_a^{\dagger}(n)\tilde{C}_a(n)\right).
\]
The operators of $Q^-_{\rm XS}$ induce extra terms in the Hamiltonian
\begin{eqnarray}
P^-_{\rm XS}&=&\frac{1}{2\sqrt{2}}\{Q^-_{\rm XS},Q^-_{\rm XS}\}\label{PmXS}\\
&=&\frac{g^2\hat{v}^2L}{2\pi}\sum_{n=1}^{\infty}\frac{1}{n}
\left(D_a^{\dagger}(n)D_a(n)+\tilde{D}_a^{\dagger}(n)\tilde{D}_a(n)
+C_a^{\dagger}(n)C_a(n)+\tilde{C}_a^{\dagger}(n)\tilde{C}_a(n)\right),
\nonumber
\end{eqnarray}
which are {\em bona fide} mass terms with
correct dimensions, since the VEV is dimensionless, and $g$ has dimension
of mass in the two-dimensional theory.
Note that there are additional induced terms, e.g.~$\{Q_{\rm SYM},Q_{\rm XS}\}$.
To obtain the spectrum of the theory one has to solve the matrix eigenvalue
problem
\beq
2P^+P^-|n\rangle=M^2_n|n\rangle,\label{EVP}
\eeq
which will yield the mass (squared) eigenvalues $M^2_n$, 
and the eigenfunctions of the
bound states of the theory, parametrized by the harmonic resolution $K$ induced
by the compactification and related to the light-cone momentum 
$P^+=\frac{\pi}{L}K$. When generating matrix elements it becomes convenient 
to use the rescaled parameters 
$v=\hat{v}/\sqrt{N_c}$ and 
$\kappa=\hat{\kappa}/\sqrt{N_c}$, 
because the effective gauge coupling is
$g\sqrt{N_c}$. We will refer to $v$ and $\kappa$ as the VEV and the Chern-Simons
coupling in the remainder of the paper.

\begin{table}
\centerline{
\begin{tabular}{|c|cc|}\hline
$\cal A$ & $\cal PA$ & $\cal OA$\\ \hline
$Q^-_{\rm SYM}$ & $+Q^-_{\rm SYM}$ & $+Q^-_{\rm SYM}$\\
$Q^-_{\rm fund}$ & $+Q^-_{\rm fund}$ & $+Q^-_{\rm fund}$\\
$Q^-_{\rm CS}$ & $-Q^-_{\rm CS}$ & $+Q^-_{\rm CS}$\\
$Q^-_{\rm XS}$ & $-Q^-_{\rm XS}$ & $-Q^-_{\rm XS}$\\\hline
\end{tabular}}
\caption{\label{TableSymmetry}
Transformation properties of the parts of the supercharge under
parity $\cal P$ and orientation reversal $\cal O$ as defined in the text.}
\end{table}

\section{Symmetries}

The theory is marginally invariant under two discrete symmetries, in the 
sense that different parts of the supercharge will respect different 
symmetries. 
Parity acts on the annihilation operators introduced in 
Eqns.~(\ref{expandA2})-(\ref{expandPsi}) as follows
\begin{equation}
{\cal P}: A_{ab}\rightarrow -A_{ab}, \,\, B_{ab}\rightarrow 
B_{ab},\quad
 C_{a}\rightarrow -C_{a}, \,\, \tilde{C}_{a}\rightarrow 
-\tilde{C}_{a},\quad
D_{a}\rightarrow D_{a}, \,\, \tilde{D}_{a}\rightarrow 
\tilde{D}_{a}.
\label{parity}
\end{equation}
Note that this is the light-cone analogue of parity, and as such 
the transformations might not be intuitively clear.
The Hamiltonian $P^-$ commutes with the parity operator 
only
in the absence of 
both the VEV-induced mass terms and the
Chern-Simons term, Eq.~(\ref{qcs}), 
the latter  
mixes parity-odd with parity-even adjoints. 
Mass eigenvalues are degenerate under parity, but not under the 
$\cal O$ symmetry \cite{Kutasov94} reversing the orientation of a string  
of partons
\beq
{\cal O}: A_{ab}\rightarrow -A_{ba}, \quad B_{ab}\rightarrow -B_{ba},
\quad C_{a}\rightarrow \tilde{C}_{a}, \quad 
D_{a}\rightarrow -\tilde{D}_{a}.
\label{Tsym}
\eeq
Adding the mass terms, ${Q}^-_{\rm XS}$, to the supercharge 
without the Chern-Simons term
destroys both symmetries
yet the combination ${\cal PO}$ is intact as inspection shows, with 
doubly degenerate eigenvalues. No symmetry remains if both Chern-Simons
and mass terms are present, although mass differences between nominal 
parity partners are extremely small.
For a summary of the symmetry properties of the parts of the supercharge, 
see Table \ref{TableSymmetry}.
Obviously, symmetry is restored as  
$\kappa\rightarrow\infty$ (${\cal O}$ is a good symmetry), and as
$v\rightarrow\infty$ (${\cal PO}$ good).

%
\section{Analytical Results}
\label{sectAnaResults}

For $K=3$ we can solve the matrix Eq.~(\ref{EVP}) in closed form
in the absence of a Chern-Simons term.
We can then use the
discrete symmetry $\cal PO$ to reduce the number of bosonic and fermionic 
states to four, say in the ${\cal PO}$ even sector, with states 
\begin{eqnarray*}
|1\rangle_{b+}&=&\frac{1}{\sqrt{2N_c}}\,\Tr\!\left[\tilde{C}^{\dagger}(2){C}^{\dagger}(1)+\tilde{C}^{\dagger}(1){C}^{\dagger}(2)]\right]|0\rangle,\\
|2\rangle_{b+}&=&\frac{1}{\sqrt{2N_c}}\,\Tr\!\left[\tilde{D}^{\dagger}(2){D}^{\dagger}(1)-\tilde{D}^{\dagger}(1){D}^{\dagger}(2)]\right]|0\rangle,\\
|3\rangle_{b+}&=&\frac{1}{{N_c}}\,\Tr\!\left[\tilde{C}^{\dagger}(1){A}^{\dagger}(1){C}^{\dagger}(1)\right]|0\rangle,\\
|4\rangle_{b+}&=&\frac{1}{\sqrt{2}N_c}\,\Tr\!\left[\tilde{C}^{\dagger}(1){B}^{\dagger}(1){D}^{\dagger}(1)+\tilde{D}^{\dagger}(1){B}^{\dagger}(1){C}^{\dagger}(1)]\right]|0\rangle,\\
|1\rangle_{f+}&=&\frac{1}{\sqrt{2N_c}}\,\Tr\!\left[\tilde{C}^{\dagger}(2){D}^{\dagger}(1)+\tilde{D}^{\dagger}(1){C}^{\dagger}(2)]\right]|0\rangle,\\
|2\rangle_{f+}&=&\frac{1}{\sqrt{2N_c}}\,\Tr\!\left[\tilde{C}^{\dagger}(1){D}^{\dagger}(2)+\tilde{D}^{\dagger}(2){C}^{\dagger}(1)]\right]|0\rangle,\\
|3\rangle_{f+}&=&\frac{1}{\sqrt{2}N_c}\,\Tr\!\left[\tilde{C}^{\dagger}(1){A}^{\dagger}(1){D}^{\dagger}(1)+\tilde{D}^{\dagger}(1)A^{\dagger}(1){C}^{\dagger}(1)]\right]|0\rangle,\\
|4\rangle_{f+}&=&\frac{1}{N_c}\,\Tr\!\left[\tilde{D}^{\dagger}(1){B}^{\dagger}(1){D}^{\dagger}(1)]\right]|0\rangle.\\
\end{eqnarray*}
In this basis the supercharge matrix reads
\[
\left({\cal Q}_{b+}\right)=\,_{f+}\!\langle m|Q^-|n\rangle_{b+}=
-\frac{ig\sqrt{N_c L}}{2^{1/4}\pi}
\left(
\begin{array}{cccc}
\sqrt{\pi}v & \sqrt{\frac{\pi}{2}}v & \frac{1}{2\sqrt{2}} & 0\\
\sqrt{\frac{\pi}{2}}v & -\sqrt{\pi}v & \frac{1}{2} & i\sqrt{2}\\
0 & \frac{1}{\sqrt{2}} & \sqrt{2\pi}v & 0 \\
i\frac{3}{2\sqrt{2}} & -i & 0 & \sqrt{2\pi}v 
\end{array}
\right).
\]
We note that $Q^-_{\rm SYM}|n\rangle_{b+}=0$ for $K=3$, and that
the eigenvalues of the two ${\cal PO}$ sectors are degenerate. 
The solutions of the 
characteristic polynomial of ${\cal Q}_{b+}({\cal Q}_{b+})^{\dagger}$
\[
\lambda^4+a_1\lambda^3+a_2\lambda^2+a_3\lambda+a_4=0,
\]
with
\begin{eqnarray*}
a_1&=&-5-7\pi v^2,\\
a_2&=&\frac{451}{64}+25\pi v^2+\frac{73}{4}\pi^2 v^4,\\
a_3&=&-\frac{255}{128}-\frac{615}{32}\pi v^2
-\frac{165}{4}\pi^2v^4-21\pi^3v^6,\\
a_4&=&9\pi^4\left(v^2-\frac{\sqrt{33}-5}{8\pi}\right)^2
\left(v^2+\frac{\sqrt{33}+5}{8\pi}\right)^2,
\end{eqnarray*}
are related to the eigenvalues $M^2_n$ of the mass (squared) operator 
in the bosonic ${\cal PO}$-even sector
\[
{\cal M}^2|_{b+}=2P^+P^-|_{b+}=
\sqrt{2}P^+{\cal Q}_{b+}({\cal Q}_{b+})^{\dagger},
\] 
by letting 
$M^2_n=\sqrt{2}\frac{\pi K}{L}
\left(\frac{g\sqrt{N_c L}}{2^{1/4}\pi}\right)^2\lambda$.

Solving this generic quartic equation is intricate. For our
purposes it suffices to determine at which values of the VEV massless 
eigenstates exist. For $\lambda=0$ to be a solution, the constant term $a_4$
has to vanish, which is obviously the case for
$v=((\sqrt{33}-5)/8\pi)^{1/2}\approx 0.172$. The theory of quartic equations
furthermore asserts that $\lambda$ becomes unity when $1+\sum_{i=1}^4 a_i=0$,
which gives rise to another quartic equation in $v^2$.


What type of spectrum do we expect for extreme choices of the parameters $v$ 
and $\kappa$? As $v\rightarrow\infty$ at fixed $\kappa$, we have 
$P^-\rightarrow P^-_{\rm XS}$, and obtain a free spectrum of states with one
fundamental parton of mass $m=\pi$ in units $g^2\hat{v}^2/\pi$ at each end of a 
chain of adjoints. Clearly, the lowest mass is, in the same units, 
\beq\label{LowestMass}
M^2_{lowest,v\rightarrow\infty}=K(\frac{\pi}{K/2}+\frac{\pi}{K/2})=4\pi, 
\eeq
and the highest mass ({\em e.g.}~of
two fundamentals with smallest possible momenta linked
by $K-2$ adjoints) is $M^2_{highest}=2\pi K$. 
We note that their degeneracies are
vastly different. There are 4(8) states of lowest mass at even(odd) $K$, but
$8\cdot 3^{K-3}$ of mass $2\pi K$. 
The former states are part of a set of $4(K-1)$ states with no adjoint partons which 
have generally the smallest masses in the spectrum. 
According to Eq.~(\ref{LowestMass}), the smallest mass is
obtained when two fundamentals split the total momentum, {\em i.e.}~when both
have large momentum. Obviously, adding more (adjoint) partons increases the mass.
The same reasoning leads to a similar free particle spectrum for large $\kappa$
at fixed $v$. 
Here, the $4(K-1)$ states with no adjoint partons will be massless, and the
lightest massive states will have a mass (squared) of $\frac{K\pi}{K-2}$ in units 
$\hat{\kappa}^2/\pi$. 

It seems clear that light states at large $v$ or $\kappa$ will remain
light as these parameters decrease. 
We thus conclude that the lighter states in the spectrum, regardless of
the values of $v$ or $\kappa$, will be the ones in which 
the {fundamental} partons have large momenta, 
{\em i.e.}~states with a minimal number of partons. 
Interaction between states of different parton number will increase the 
average number of partons in these states, but not dramatically.

%
\section{Numerical Results}
\label{sectNumResults}

\subsection{Masses as functions of the physical parameters}

\begin{figure}
\centerline{
\psfig{file=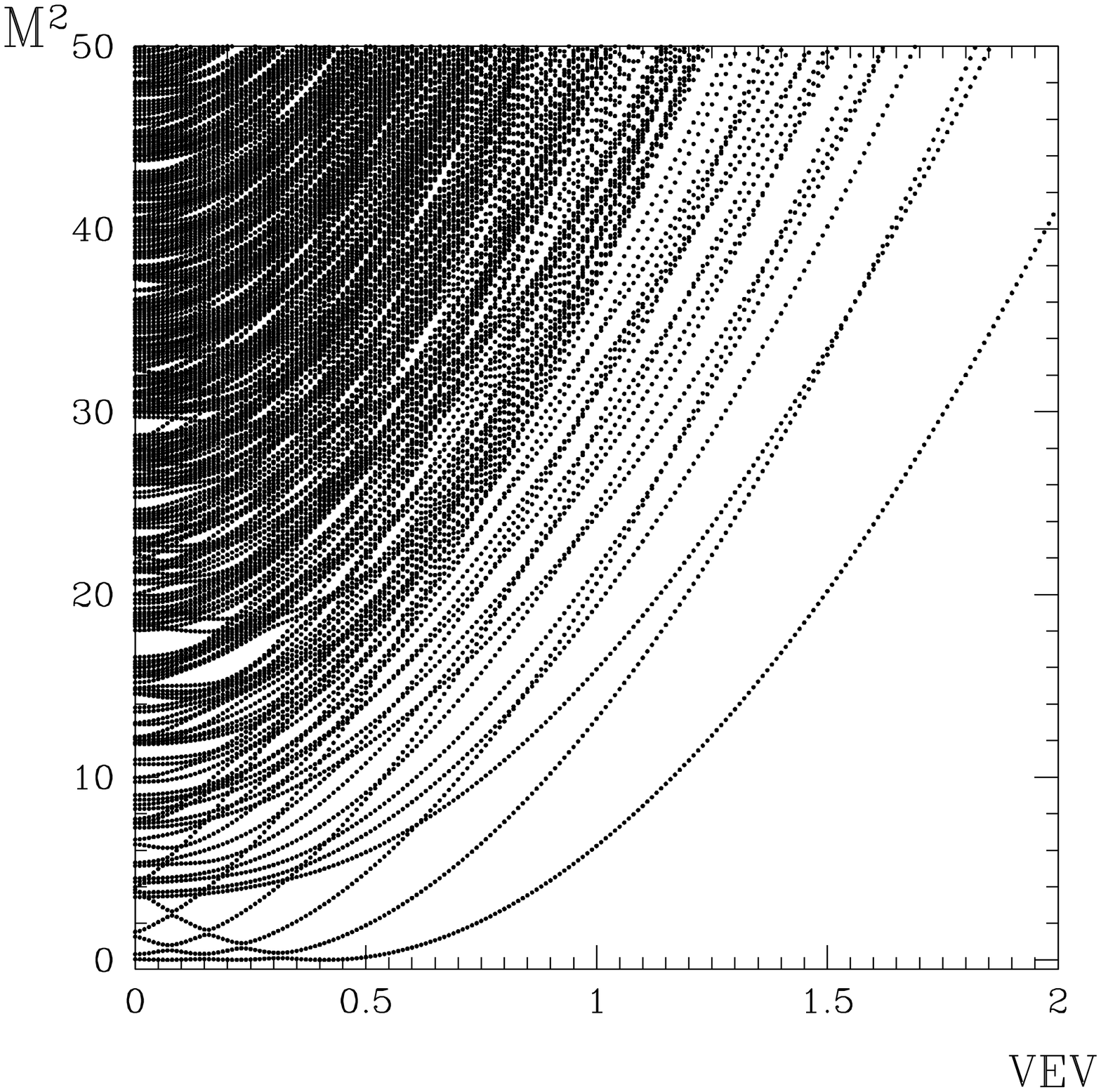,width=8cm}
\psfig{file=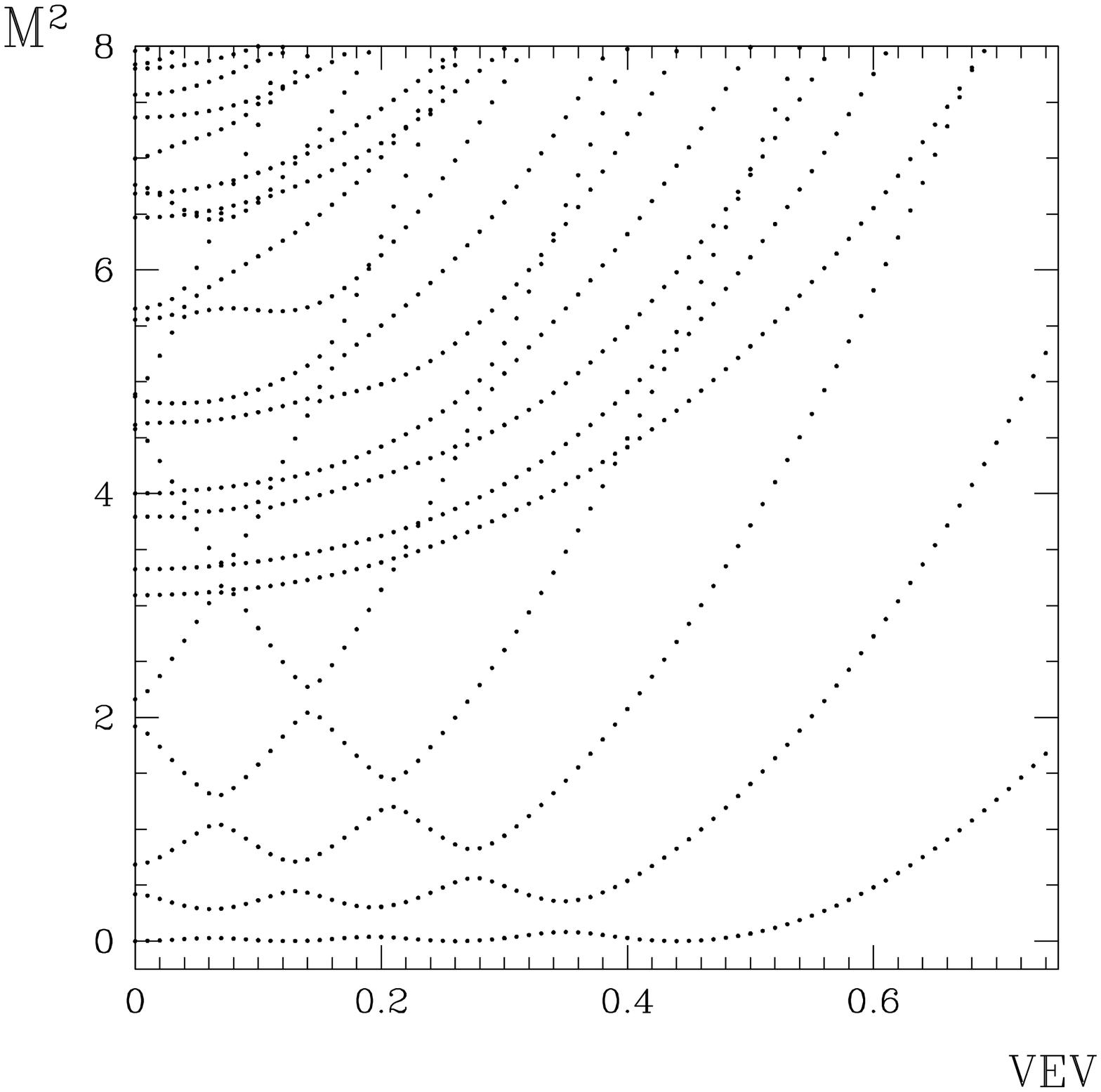,width=8cm}}
\caption{\label{specVEV7}The 
spectrum as a function of the VEV at $\kappa=0$ 
in the bosonic $\cal PO$-even sector: 
(a) overview at $K=7$ (left); and 
(b) detailed view of the lightest states at $K=8$ (right). Masses (squared) 
are in units $g^2N_c/\pi$.}
\end{figure}

We plot the spectrum of the theory without CS term as a function of the VEV, 
Fig.~\ref{specVEV7}, 
and see that the most prominent feature is a quadratic 
rise of masses (squared) with $v$.
At small $v$ we find $4(K-1)$ light states, four of which (one in each 
symmetry sector) 
become exactly\footnote{
We showed in Sec.~\ref{sectAnaResults} analytically that one of the states in each 
sector becomes massless for a certain value of $v$. 
Judging from numerical evidence, we can assume that 
this remains true at higher $K$.}
massless at various values of $v$. 
How do bound-state masses decrease as their constituents increase in mass?
The change of mass at small $v$ is roughly\footnote{Since at $K=3$ we have
decreasing masses but $Q^-_{\rm SYM}=0$, $Q^-_{\rm XS}$ 
alone is responsible for the decrease; likely also at higher $K$.}
$
\delta M^2_n = \langle n |2P^+P^-_{\rm XS}| n\rangle.
$ 
It is clear from the form of the Hamiltonian, Eq.~(\ref{PmXS}), 
that a decrease in bound-state mass is possible even
as $v$ grows for states having 
overlap with certain basis states containing two fermionic fundamentals
due to fermionic statistics. 
As we saw in Sec.~\ref{sectAnaResults}, the lightest states will be short, 
having a large overlap with these special basis states. In fact, this seems 
to be 
the very reason why they are light. Close inspection of Fig.~\ref{specVEV7}
shows that a few of the more massive states see their mass 
decreasing over some $v$ range as well, which is, of course, not contradicting
our finding.

%
%

\begin{figure}
\centerline{
\psfig{file=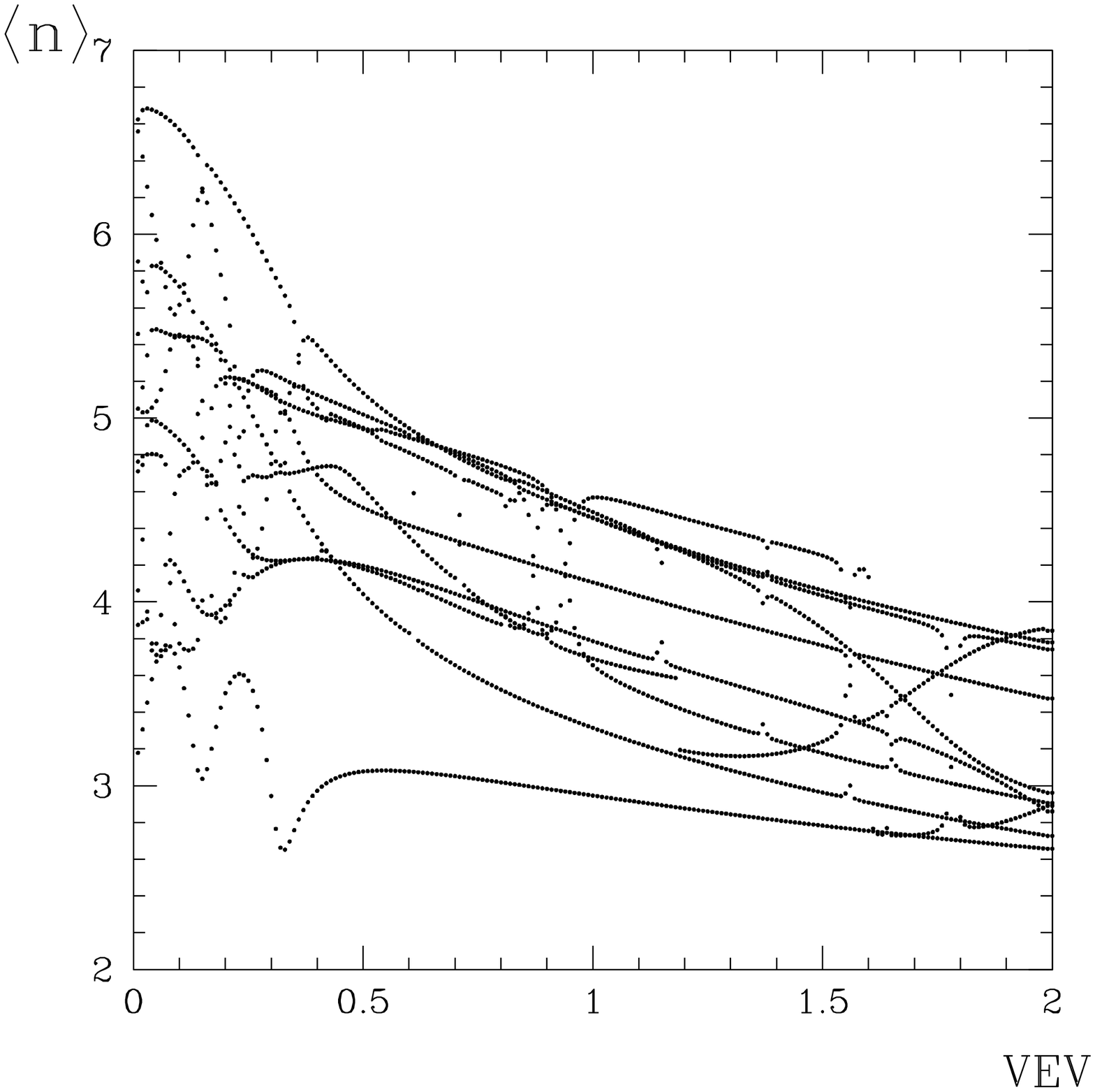,width=8cm}
\psfig{file=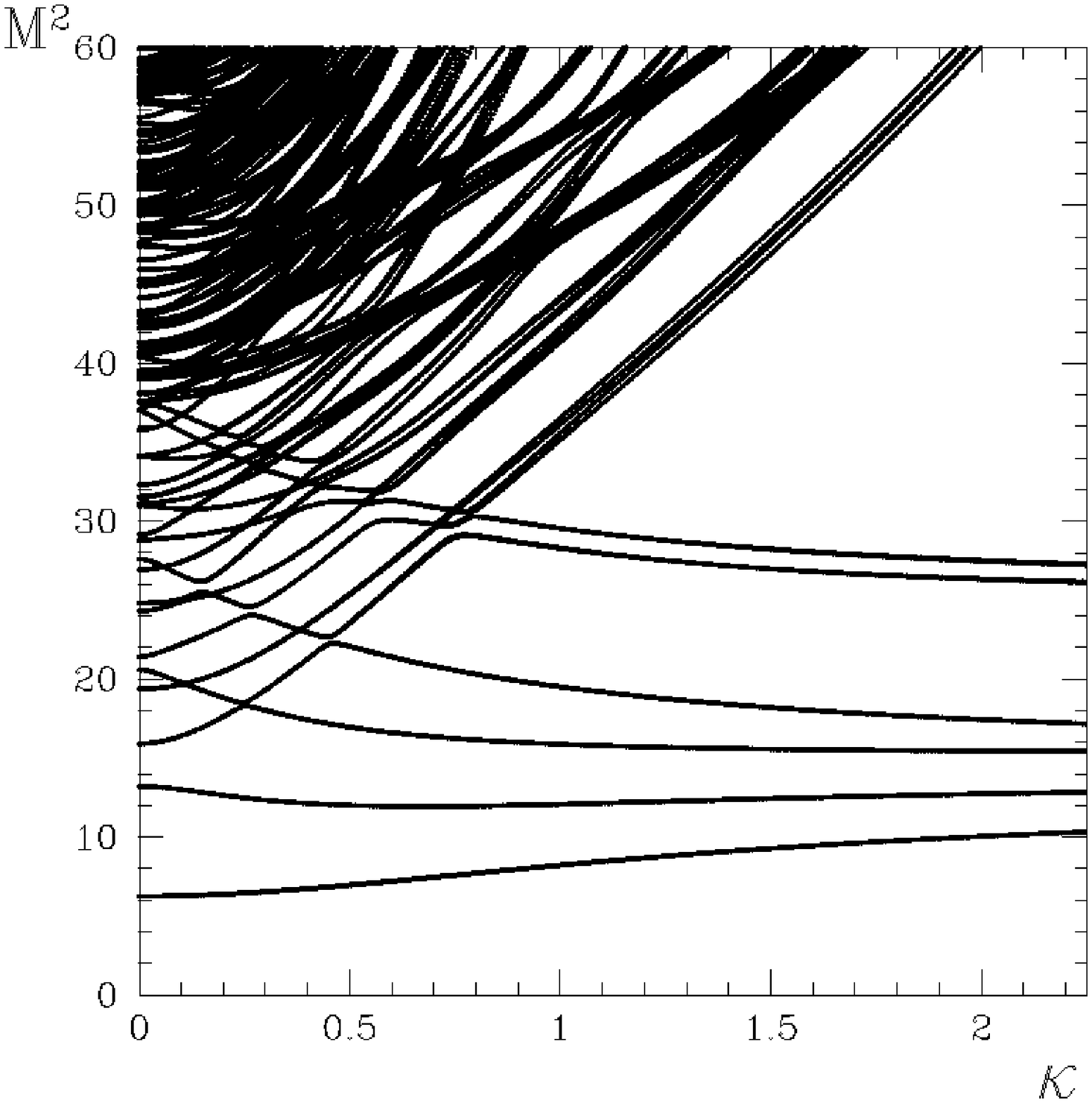,width=8cm}}
\caption{\label{avgn} Figure showing (a) 
the average parton number of the ten lightest states 
as a function of the VEV at at $\kappa=0$ and $K=7$ (right); and
(b) the bosonic 
spectrum with masses (squared) in units $g^2N_c/\pi$ 
as a function of the Chern-Simons coupling 
${\kappa}$ in units $g$ at $v=1$ and $K=7$.}
\end{figure}

From the analytical considerations in Sec.~\ref{sectAnaResults}
it is clear that the lowest $4(K-1)$ states
are special in that they do not contain adjoint partons at large $v$. 
The quadratic nature of the interaction terms suggests
to interpret the pattern of masses $M^2_n(v)$ in Fig.~\ref{specVEV7}
as an overlay of $K-1$ parabolas
with centers shifted in $v$, and individual 
masses distorted by eigenvalue repulsion. 
Indeed, for even (odd) $K$ we see that at
$\frac{K}{2}$ ($\frac{K-1}{2}$) VEV values a massless state is present.
At fixed $K$, all masses eventually rise with $v$, because in the tug-o-war 
between the VEV and the effects of the admixture of fermionic ``mass-reducing''
basis states, the former must win. On the other hand, if $K$ grows, so does 
the number of these basis states, keeping the effects of the 
growing VEV in check over a larger region.  
As more states come on line at higher $K$, the 
parabolas will add up to 
straighter lines at lower and lower mass, and
we speculate that in the continuum limit 
a countable-infinite number of massless states emerges, 
all being linear combinations
of infinitely many two-parton states. This hypothesis is supported by 
the behavior of average number of partons $\langle n\rangle$
of the lowest states. We saw in Sec.~\ref{sectAnaResults}
that $\langle n\rangle=2$ for $v\rightarrow\infty$. 
Moreover, Fig.~\ref{avgn}(a), displaying the average parton numbers of the 
ten lightest states at $K=7$ as a function of the VEV,
shows a trend towards short states already for intermediate values of $v$. 
It seems thus that at large $K$ we could have a separation between (almost)
massless states and heavy states. We come back to this issue in the next 
subsection.

%
%


The effect of adding a Chern-Simons term on the spectrum is as anticipated, see 
Fig.~\ref{avgn}(b). 
The adjoints become massive, and the only states remaining light as $\kappa$ 
grows are the $4(K-1)$ two-parton states without adjoints.
Note that each ``line'' $M_n^2(\kappa)$ in the plot is actually a double line
of two almost degenerate mass eigenvalues.
The reason for the approximate degeneracy is that the states (at 
least the light ones) are 
largely devoid of partons subject to the symmetry-breaking CS term.
Remarkably, the symmetry-breaking is very small for all states.
The main effect of the CS term is to lift the masses of all bound-states.
As $\kappa$ grows, the spectrum at fixed resolution $K$ 
separates into light and very heavy states.
%
\begin{figure}
\centerline{
\psfig{file=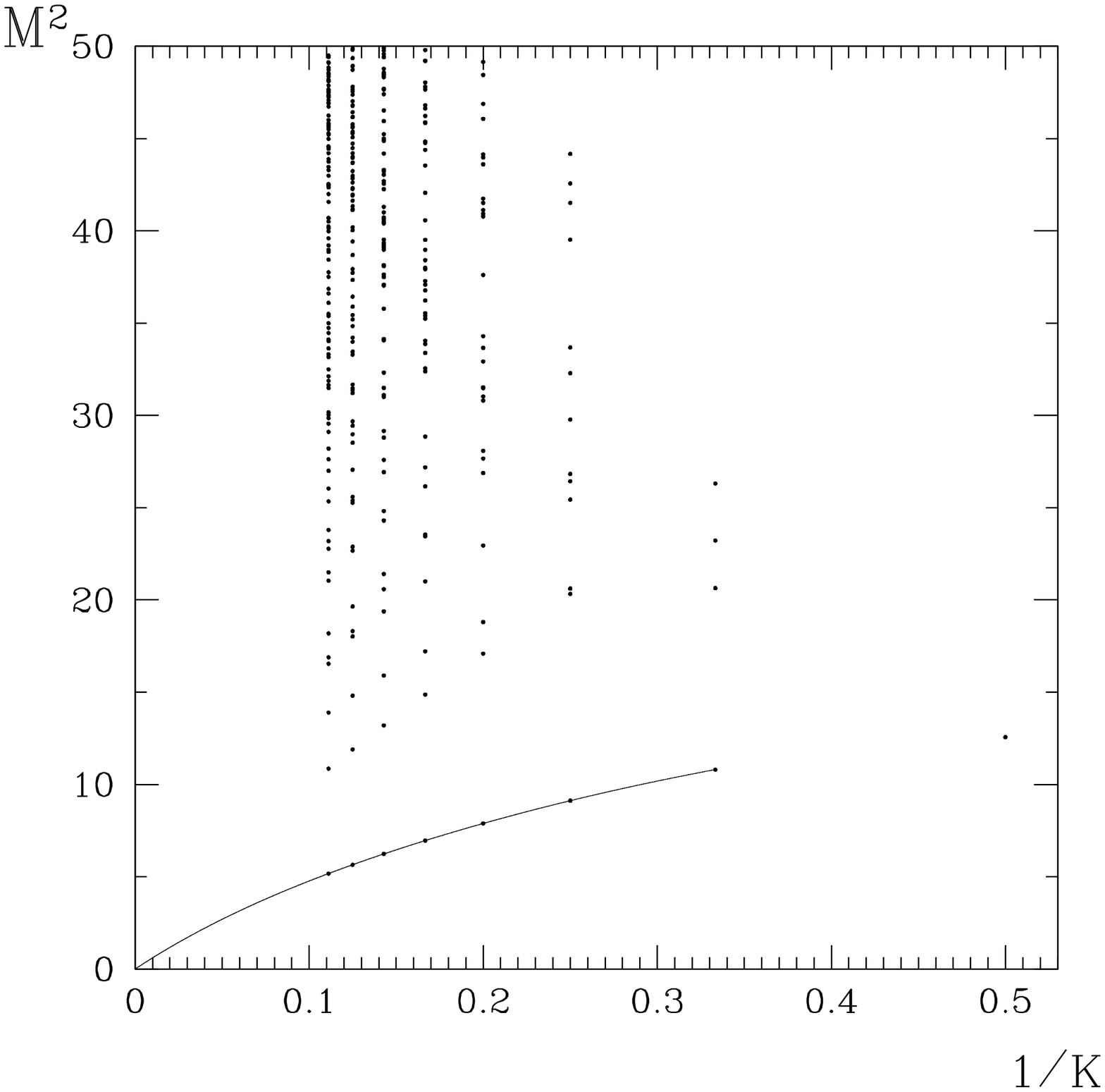,width=8cm}
\psfig{file=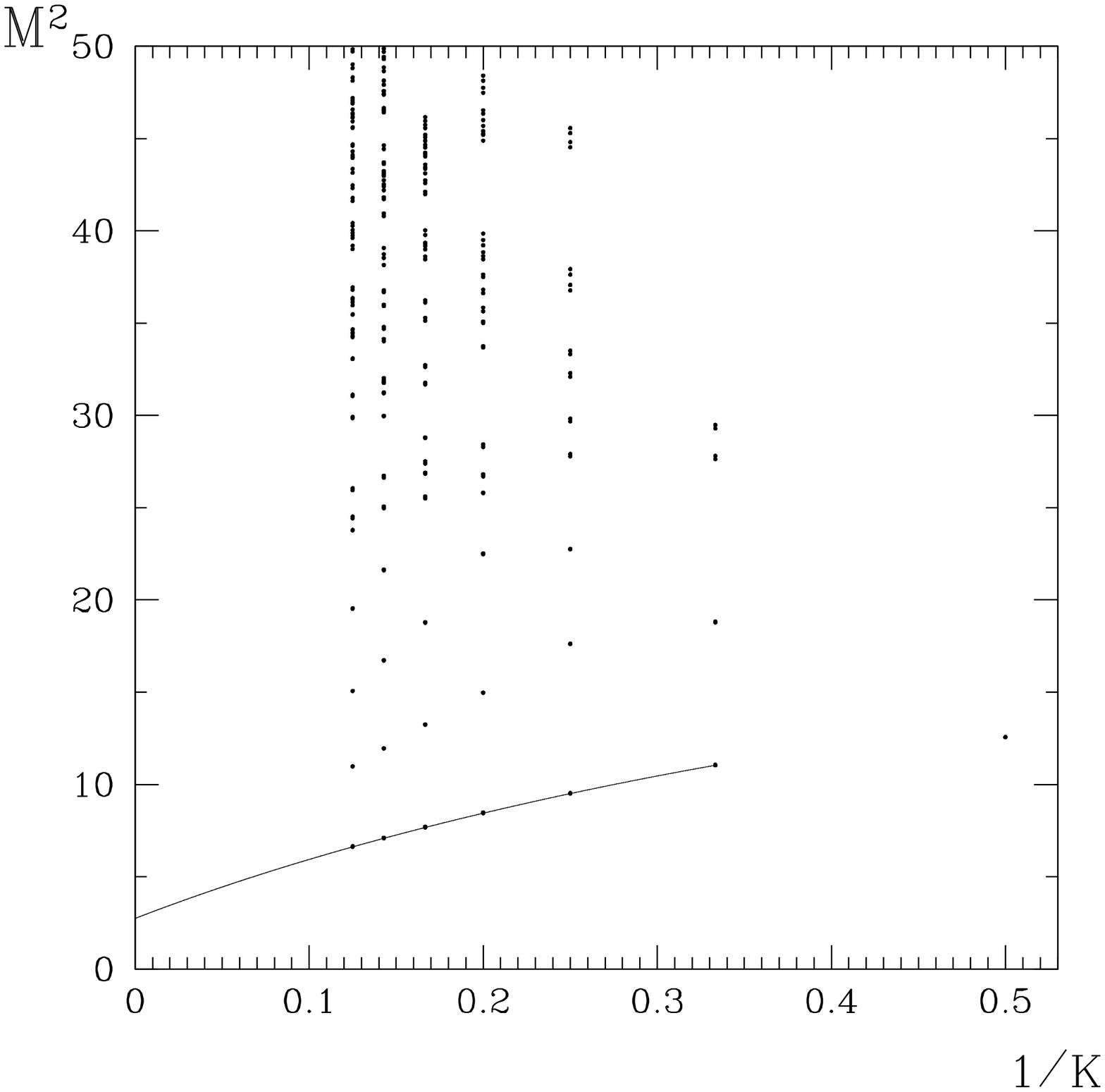,width=8cm}}
\caption{\label{spectrumK} Figure showing (a) the bosonic 
spectrum as a function of the inverse harmonic resolution:
at $v=1$ (left); and 
(b) The spectrum at ${\kappa}=g/\sqrt{\pi}$, $v=1$ as a function
of the inverse resolution (right). Masses (squared) are in 
units $g^2N_c/\pi$.}
\end{figure}
%

\subsection{The continuum limit}

While we are able to present evidence that an infinite set of massless 
states
exists for all values of the VEV, we can demonstrate decisively that a massless
state exists at a specific VEV. We do this by plotting the masses
as a function of the inverse harmonic resolution $1/K$ at fixed VEV, say $v=1$,
in Fig.~\ref{spectrumK}(a) and extrapolating to the continuum limit. 
%
%
A fit of seven
data points to a polynomial of fifth degree, included in 
Fig.~\ref{spectrumK}(a), yields a continuum mass (squared) of
$
M^2_{lowest,v=1,\kappa=0}(K\rightarrow\infty)=0.0015 \pm{0.2257}
$ 
in units $g^2N_c/\pi$.
We attributed a systematic error to this value by performing a fit to 
a polynomial of fourth and sixth degree, respectively, and
taking the larger difference between these extrapolations and the
value above as the
uncertainty. The continuum mass is thus consistent with zero.

On the other hand, there is no massless state in the continuum limit if a 
Chern-Simons term is present. Plotting the spectrum at $v=1$ and 
${\kappa}=g/\sqrt{\pi}$ as a function of $1/K$ 
in Fig.~\ref{spectrumK}(b), we see that a
fit to a polynomial of fourth degree in $1/K$ to 
the masses of the lightest states (six data points)
suggests that no massless states exist when $\kappa\neq 0$, since
$
M^2_{lowest,v=1,\kappa=1}(K\rightarrow\infty)=2.74\pm 0.30$
in units $g^2{N_c/\pi}$, 
where we estimated the systematic error as described above.

\begin{figure}
\centerline{
\psfig{file=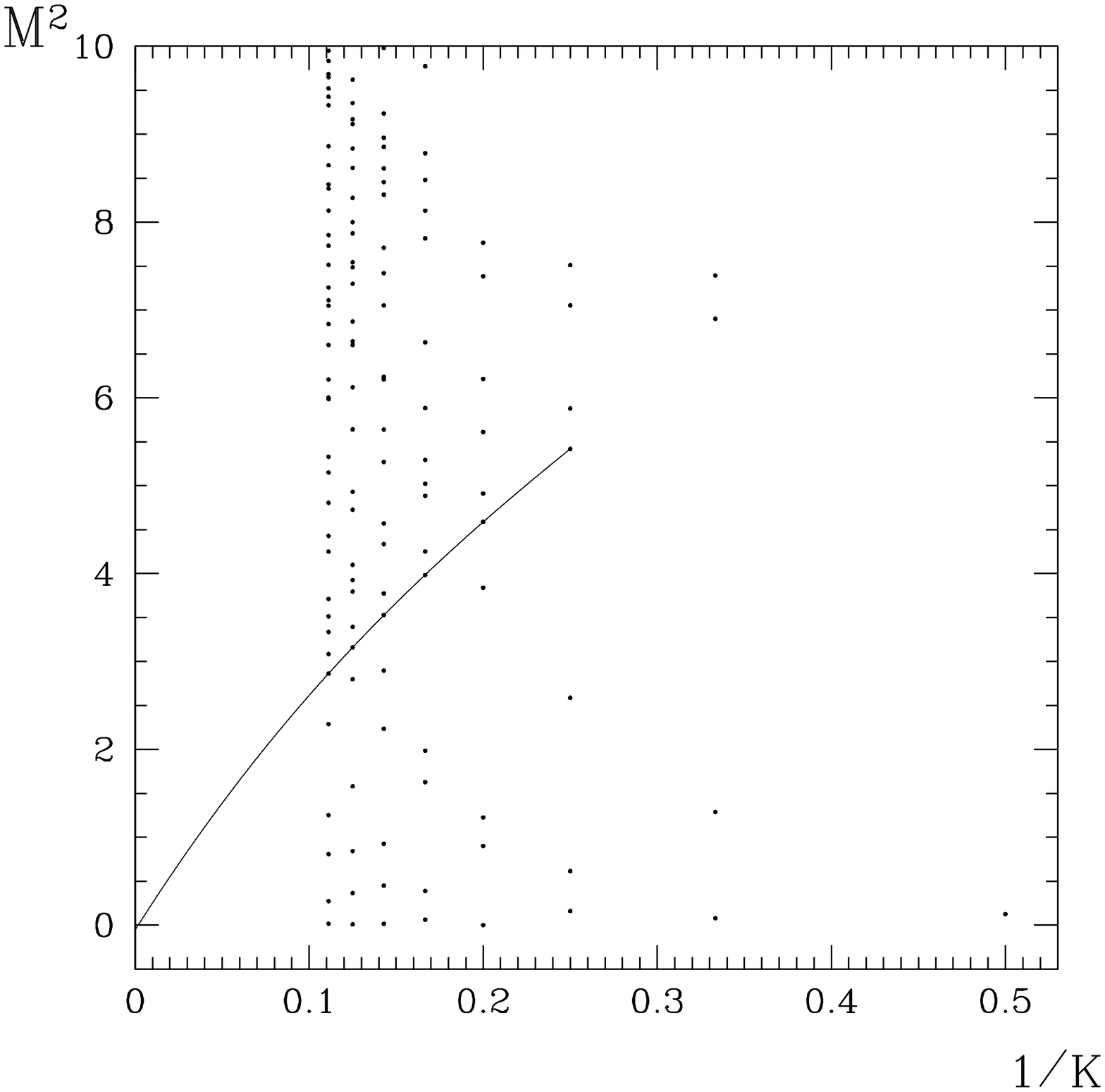,width=8cm}
\psfig{file=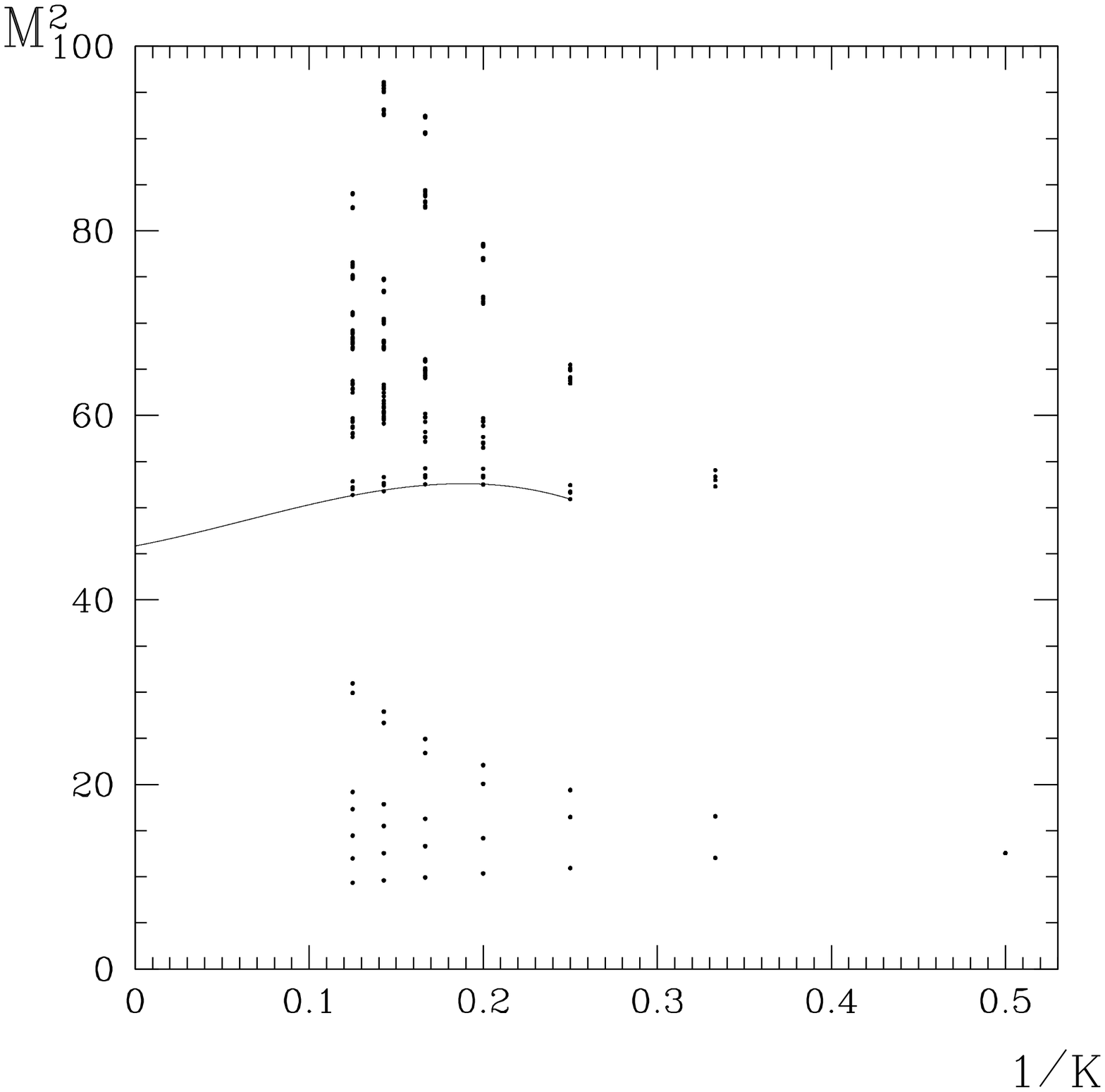,width=8cm}}
\caption{\label{spectrumK2}The spectrum  as a function
of the inverse resolution: (a) at $v=0.1$, $\kappa=0$; and 
(b) at $v=1$, ${\kappa}=3g/\sqrt{\pi}$. 
Masses (squared) are in units $g^2N_c/\pi$.}
\end{figure}

Finally, we need to settle the question whether a prominent feature of the spectra
as a function of the physical parameters, namely the gap between light and heavy
states, persists as the unphysical parameter $K$ is removed in 
the continuum limit. Figures~\ref{spectrumK2}(a) and (b) reveal that the gap
survives the limit if the Chern-Simons coupling is substantial, whereas
in its absence, the gap seems to collapse, at least at small $v$. Both fitting 
functions are polynomials of third degree. The estimated errors are large
enough at $\kappa=0$ to prevent us from concluding that the heavier states
remain massive. We were unable to perform the analysis at substantially
different parameters because we could not label states unambiguously, {\em i.e.}
decide whether they belong to the light or the heavy 
states.
 
%
%

\subsection{Structure functions}

It is interesting to look at the wavefunctions of the bound-states.
In the full theory (massive fundamental and adjoint partons) the amount 
of information encoded in them is too large to be useful. Instead, we display 
the expectation values of various particle number operators in the lightest two
states in Table \ref{TableAVGN} as a function of $K$.
It is obvious that symmetry breaking is exceedingly small and that it 
does not grow
significantly with $K$. The two lightest states have almost identical 
properties. The fact that the average parton number is creeping up as $K$ 
grows is caused by the increase of the number of adjoint bosons in the
states. Both average fundamental boson and fermion number are roughly one,
implying that the state consists largely of two-parton basis states, roughly 
half of which have two bosons, the other half having two fermions. 
Some of the latter have the effect of 
lowering the bound-state mass, producing the light mass we observe. 

\begin{figure}
\centerline{
\psfig{file=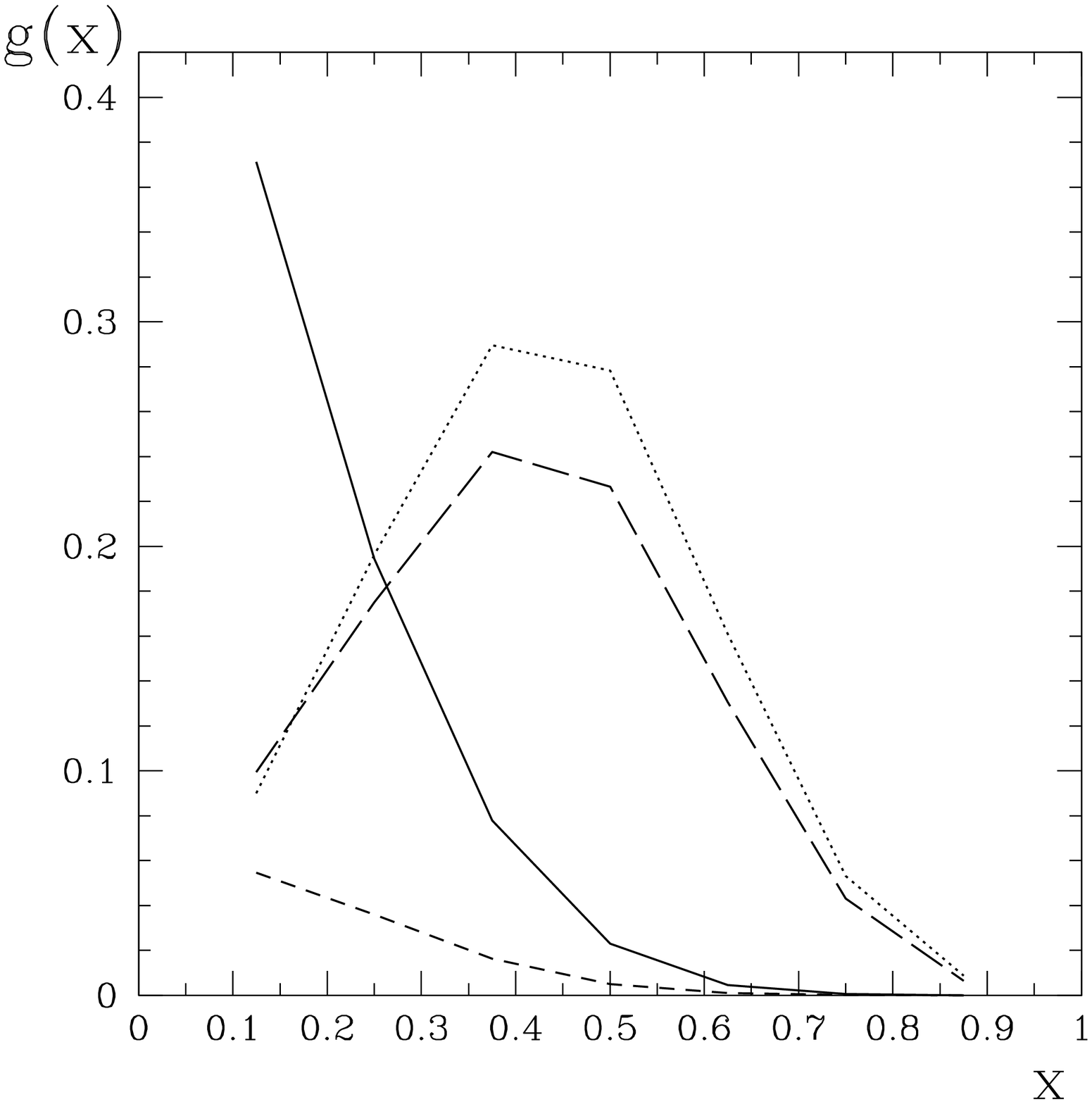,width=8cm}
\psfig{file=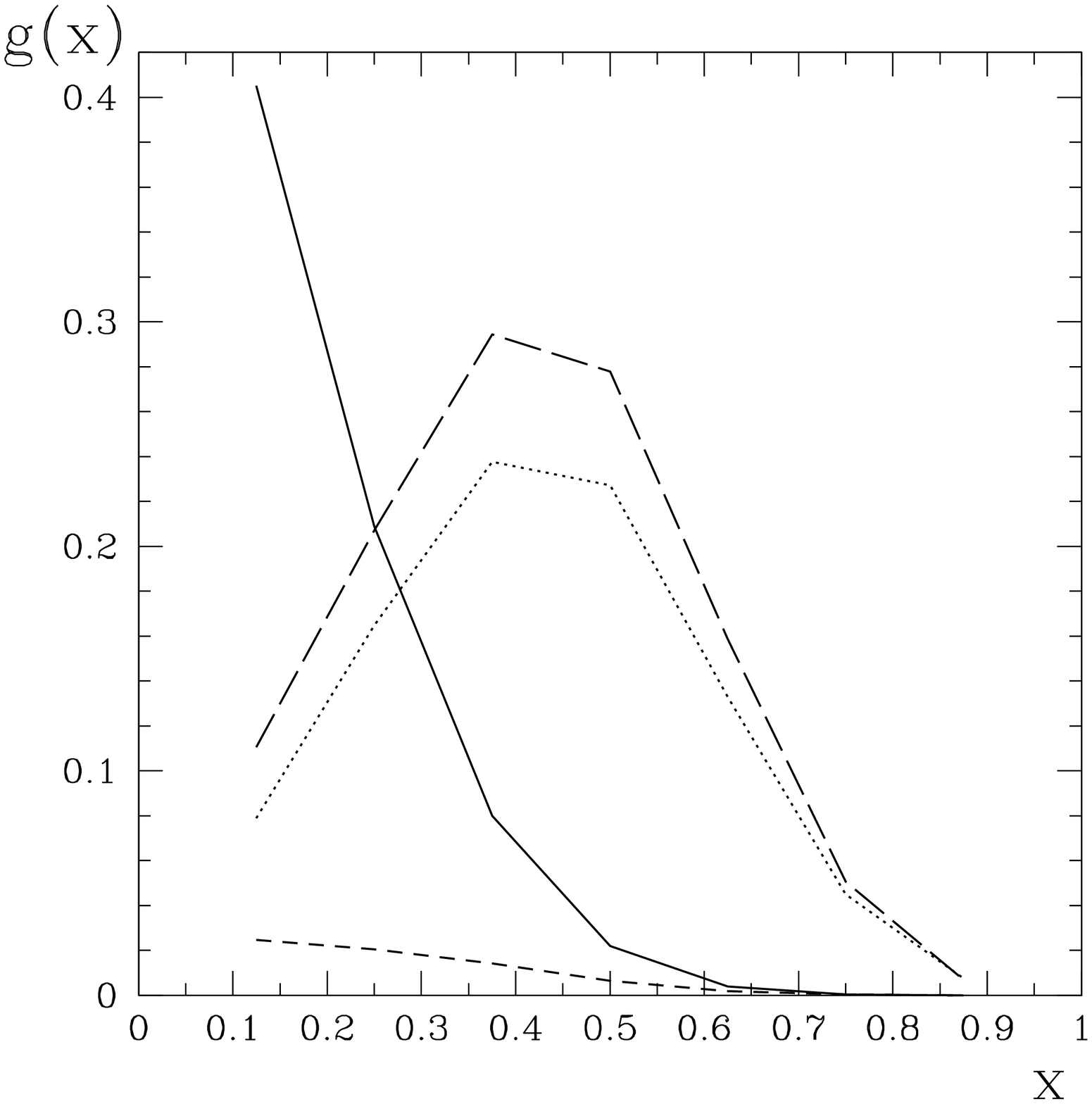,width=8cm}}
\caption{\label{SFs}Structure functions of the lowest two bosonic states 
as a function of the longitudinal momentum fraction $x=n/K$ at 
$v=1$, $\kappa=g/\sqrt{\pi}$ and $K=8$: 
(a) lowest state; and (b) second lowest state. 
Solid lines: $g_{aB}$. Long-dashed lines: $g_{fF}$. 
Short-dashed lines: $g_{aF}$. Dotted lines: $g_{fB}$.}
\end{figure}

We define the discrete version  
of the customary structure functions at harmonic resolution $K$ as
\[
g_A(n)=\sum^K_{q=2}\sum^{K-q}_{n_1,\ldots,n_q=1}
\delta\!\left(\sum_{i=1}^q n_i-K\right) \sum_{l=1}^q \delta_n^{n_l}
\delta^A_{A_l}|\psi(n_1,\ldots,n_q)|^2.
\]
They are normalized such that the summation over the argument ($n$) yields the
average number of type $A$ partons in a state. 
The possible types are adjoint bosons (aB), 
adjoint fermions (aF), fundamental bosons (fB), and 
fundamental fermions (fF). 
What do we expect? From Sec.~\ref{sectAnaResults} we know that the lightest 
states are short, and to minimize mass, their fundamental partons should gobble
up as much of the total momentum as possible while splitting it evenly. 
Therefore, we anticipate the structure 
functions to be peaked around longitudinal momentum fraction $x=0.5$ for 
fundamentals, and around $x=1/K$ for adjoints. Furthermore, odds are
that there is only one adjoint parton, and we have a preference for states
with two fundamental fermions since they can lower the bound-state mass. 
Hence, in the bosonic sector, we should 
find more adjoint bosons than fermions in the light bound states.
We see from
Fig.~\ref{SFs} that our expectations are largely met for the lightest two 
states.  
Obviously, the two lightest states are very similar in mass due to the small
breaking of the $\cal PO$ symmetry, and they also have very similar 
eigenfunctions as evident from their structure functions. Apart from the flip
in importance of $g_{fF}$ with $g_{fB}$, there is a minor reduction of $g_{aF}$
towards smaller $x$ in the heavier state. The lighter state has 
a slightly smaller number of fundamental fermions which is somewhat surprising.
Since only some of the two-fundamental-fermion basis states can lower the 
bound-state mass, this 
is not in conflict with our previous conclusions, however.

\begin{table}
\centerline{
\begin{tabular}{|c|rccccc|}
\hline
$K$ & $M^2$ & $\langle n\rangle$ &$\langle n_{aB}\rangle$ &
$\langle n_{fB}\rangle$ &$\langle n_{aF}\rangle$ &$\langle n_{fF}\rangle$ \\ 
\hline
3 &11.0413&2.2630&0.2260&1.0045&0.0370&0.9955\\
  &11.0686&2.2627&0.2486&0.9862&0.0140&1.0138\\
4 &9.5070&2.4420&0.3803&1.0118&0.0617&0.9882\\
  &9.5413&2.4429&0.4105&0.9709&0.0324&1.0291\\
5 &8.4447 &2.5676 &0.4884 &1.0340 &0.0792 &0.9660\\
  &8.4814 &2.5699 &0.5262 &0.9439 &0.0436 &1.0561\\
6 &7.6705 &2.6595 &0.5669&1.0506&0.0926&0.9494\\
  &7.7075&2.6626&0.6097&0.9242&0.0529&1.0758\\
7 &7.0823&2.7294&0.6258&1.0646&0.1036&0.9354\\
  &7.1188&2.7330&0.6721&0.9081&0.0609&1.0919\\
8 &6.6194 &2.7844 &0.6715 &1.0771 &0.1129 &0.9230\\
  &6.6550 &2.7882 &0.7203 &0.8942 &0.0680 &1.1058\\\hline
\end{tabular}}
\caption{\label{TableAVGN}Properties of the two lightest bosonic bound states 
at $v=1$ and $\kappa=g/\sqrt{\pi}$. 
Listed are the average numbers of adjoint bosons (aB), 
fundamental bosons (fB), adjoint (aF) and fundamental fermions (fF).}
\end{table}

%
%

\section{Discussion}

We generated mass terms for the fundamental fields in a two-dimensional
SYM theory by adding a vacuum expectation 
value to the perpendicular gauge boson left over from a dimensional reduction 
of the associated three-dimensional theory.  
Supersymmetry stays intact this way. In earlier work we had fabricated an
analogous mass term for the adjoints of the theory by adding a
Chern-Simons term.  As a function of the new parameter 
$v:=\langle A^{\perp}\rangle$,
the majority of the states will see their masses increase with $v$. The
lightest state, however, experiences masslessness for several
values of $v$. We presented evidence for our hypothesis that in the 
continuum limit an infinite number of massless states will be present in this
theory for a large range of $v$ values.
As the main difference between the theories with and without a 
Chern-Simons term, we showed that in the latter case the lightest state
is massless, whereas no massless state exists if $\kappa$ is substantial.
Although we did not push for extreme numerical precision, we feel that
we have a safe handle on the continuum limit, 
meaning that the properties of individual states 
show little variation with the harmonic resolution. 

In the sense that we were able to add mass terms to all species of this 
generic  supersymmetric gauge theory, and to study their 
effects on the spectrum, this work concludes our exploration of 
two-dimensional SYM.
It would be interesting to extend our investigations to higher dimensions, 
but the 
computational effort likely will be prohibitively high if we want to strive for
the same quality in terms of convergence control.

\section*{Acknowledgments}

U.T.~would like to thank the Ohio State University's 
physics department for hospitality while this
work was completed. This work was funded partially by a grant from the 
Research Corporation.


\begin{appendix}

\section{The SDLCQ supercharge in mode decomposition}
\label{SuperchargeModes}
For completeness we list the four parts of the supercharge 
\[
Q^-=Q^-_{\rm SYM}+Q^-_{\rm fund}+Q^-_{\rm CS}+Q^-_{\rm XS}.
\]
Note that the supercharge is a Hermitian operator. Its parts are
%
\begin{eqnarray}
\label{Qminus}
Q^-_{\rm SYM}&=&-\frac{{i} g\sqrt{L}}{2^{1/4}\pi}\sum_{n_1,n_2,n_3=1}^{\infty} 
\delta_{n_1+n_2,n_3} \left\{ \frac{}{} \right.\\
&&+{1 \over 2\sqrt{n_1 n_2}} {n_1-n_2 \over n_3}
[A_{ac}^\dagger(n_1) A_{cb}^\dagger(n_2) B_{ab}(n_3)
-B_{ab}^\dagger(n_3)A_{ac}(n_1) A_{cb}(n_2) ]\nonumber\\
&&+{1 \over 2\sqrt{n_1 n_3}} {n_1+n_3 \over n_2}
[A_{ab}^\dagger(n_3) A_{ac}(n_1) B_{cb}(n_2)
-A_{ac}^\dagger(n_1) B_{ca}^\dagger(n_2)A_{ab}(n_3) ]\nonumber\\
&&+{1 \over 2\sqrt{n_2 n_3}} {n_2+n_3 \over n_1}
[B_{ac}^\dagger(n_1) A_{cb}^\dagger(n_2) A_{ab}(n_3)
-A_{ab}^\dagger(n_3)B_{ac}(n_1) A_{cb}(n_2) ]\nonumber\\
&&+({ 1\over n_1}+{1 \over n_2}-{1\over n_3})
[B_{ac}^\dagger(n_1) B_{cb}^\dagger(n_2) B_{ab}(n_3)
+B_{ab}^\dagger(n_3) B_{ac}(n_1) B_{cb}(n_2)]  \left. \frac{}{}\right\}.
\nonumber
\end{eqnarray}
\begin{eqnarray}
\label{Qfmattexp}
Q^{-}_{\rm fund}&=&-\frac{i g\sqrt{L}}{2^{1/4}\pi}
\sum_{n_1,n_2,n_3=1}^{\infty}\left\{
\frac{(n_2+n_3)}{n_1\sqrt{2 n_2 n_3}}
\left(\tilde{C}^\dagger_i(n_3)\tilde{C}_{j}(n_2)B_{ji}(n_1)\right.\right.
\nonumber\\
&&\left.-\tilde{C}^\dagger_a(n_2)B_{ab}^\dagger(n_1)\tilde{C}_{b}(n_3)
+B_{ba}^\dagger(n_1) C^\dagger_a(n_2)C_{b}(n_3)
-C^\dagger_a(n_3)B_{ab}(n_1)C_{b}(n_2) \right)\nonumber\\
&&+\frac{1}{n_1}
\left( \tilde{D}^\dagger_a(n_2)B^\dagger_{ab}(n_1)\tilde{D}_b(n_3)+ \tilde{
D}^\dagger_a(n_3)\tilde{D}_b(n_2)B_{ba}(n_1)
+B^\dagger_{ab}(n_1)D^\dagger_b(n_2) D_a(n_3)\right.
\nonumber\\
&&+
\left.D^\dagger_a(n_3) B_{ab}(n_1) D_b(n_2) \right)-\frac{i}{2\sqrt{n_2 n_3}}  
\left(C^\dagger_{a}(n_3)A_{ab}(n_2)D_{b}(n_1)\right.\nonumber\\
&&+
\left. A^{\dag}_{ab}(n_2)D^{\dag}_{b}(n_1)C_{a}(n_3)
+\tilde{D}^{\dag}_{b}(n_1)A^{\dag}_{ba}(n_2)\tilde{C}_{a}(n_3)
+\tilde{C}^{\dag}_{a}(n_3)\tilde{D}_{b}(n_1)A_{ba}(n_2)\right)\nonumber\\
&&-\frac{i}{2\sqrt{n_1 n_2}} \left(
A^\dag_{ba}(n_2)C^{\dag}_{a}(n_1)D_b(n_3)+
D^\dagger_{b}(n_3)A_{ba}(n_2)C_{a}(n_1)\right.\nonumber\\
&&\left.\left.+\tilde{D}^{\dag}_{b}(n_3)\tilde{C}_{a}(n_1)A_{ab}(n_2)
+\tilde{C}^{\dag}_{a}(n_1)A^{\dag}_{ab}(n_2)\tilde{D}_{b}(n_3)\right)
\right\}\,\delta_{n_3,n_1+n_2}.
\end{eqnarray}
The Chern-Simons term is
\begin{equation} 
\label{qcs}
%
Q^-_{\rm CS}=-\frac{i g\sqrt{L}}{2^{1/4}\pi}\left(i\sqrt{\pi}
\frac{\hat{\kappa}}{g}\right)
\sum_{n=1}^{\infty}\frac{1}{\sqrt{n}}
\left(A_{ab}^{\dagger}(n)B_{ab}(n)+B_{ab}^{\dagger}(n)A_{ab}(n)\right)\,.
\end{equation}
Finally, the extra terms induced by shifting the gauge field by its VEV are
\begin{eqnarray}
Q^-_{\rm XS}&=&-\frac{ig\sqrt{L}}{2^{1/4}\pi}\left(-i\sqrt{\pi}\hat{v}\right)
\sum_{n=1}^{\infty}\frac{1}{\sqrt{n}}\nonumber\\
&&\times\left(C_a^{\dagger}(n)D_a(n)+\tilde{C}_a^{\dagger}(n)\tilde{D}_a(n)+
D_a^{\dagger}(n)C_a(n)+\tilde{D}_a^{\dagger}(n)\tilde{C}_a(n)\right).
\end{eqnarray}
The common factor $\frac{g\sqrt{L}}{2^{1/4}\pi}$ is dropped in numerical 
calculations to obtain dimensionless matrix elements. In the mass squared operator
${\cal M}^2=2P^+P^-$ the compactification length $L$ cancels, due 
to $P^+=K\pi/L$. Its 
eigenvalues carry units of $g^2N_c/\pi$, since $N_c$ creeps in via the parton
number changing interactions, and is absorbed in a rescaling of the VEV and
Chern-Simons couplings in two-body operators: $\hat{v}=v\sqrt{N_c}$, 
$\hat{\kappa}=\kappa\sqrt{N_c}$.

\end{appendix}


\begin{thebibliography}{99}
\bibitem{SUSY2}
  K.~A.~Olive, S.~Rudaz and M.~A.~Shifman,
  ``Susy 30.'' Proceedings of the 
  International Symposium Celebrating 30 Years Of
  Supersymmetry, North-Holland, Amsterdam 2001.
\bibitem{SUSY}
  S.~P.~Martin,
  ``A Supersymmetry Primer,''
  arXiv:hep-ph/9709356.
\bibitem{DLCQ}
  S.~J.~Brodsky, H.~C.~Pauli and S.~S.~Pinsky,
  Phys.~Rept.~ {\bf 301} (1998) 299.
\bibitem{SDLCQ}
  O.~Lunin and S.~Pinsky,
  AIP Conf.~Proc.~ {\bf 494}(1999) 140.
\bibitem{Kaplan}
  D.~B.~Kaplan,
  Eur.~Phys.~J.~ST {\bf 152} (2007) 89.
\bibitem{Sakai}
  Y.~Matsumura, N.~Sakai and T.~Sakai,
  Phys.~Rev.~ D {\bf 52} (1995) 2446.
\bibitem{ImprovedSakai}
  J.~R.~Hiller, M.~Harada, S.~S.~Pinsky, N.~Salwen and U.~Trittmann,
  Phys.~Rev.~ D {\bf 71} (2005) 085008.
\bibitem{Lunin}
  O.~Lunin and S.~Pinsky,
  Phys.\ Rev.\  D {\bf 63} (2001) 045019. 
\bibitem{FundMatter2D}
  J.~R.~Hiller, S.~S.~Pinsky and U.~Trittmann,
  Phys.~Rev.~ D {\bf 67} (2003) 115005,
  Nucl.~Phys.~ B {\bf 661} (2003) 99.
\bibitem{CS2D}
  J.~R.~Hiller, S.~S.~Pinsky and U.~Trittmann,
  Phys.~Rev.~Lett.~ {\bf 89} (2002) 181602,
  Phys.~Rev.~ D {\bf 65} (2002) 085046.
\bibitem{Haney}
  P.~Haney, J.~R.~Hiller, O.~Lunin, S.~Pinsky and U.~Trittmann,
  Phys.\ Rev.\  D {\bf 62} (2000) 075002.
\bibitem{CS3D}
  J.~R.~Hiller, S.~S.~Pinsky and U.~Trittmann,
  Phys.~Rev.~ D {\bf 66} (2002) 125015,
  Phys.~Lett.~ B {\bf 541} (2002) 396.
\bibitem{Myers} R.~C.~Myers, private communication; see also
  H.~Nastase,
  arXiv:hep-th/0305069.
\bibitem{TDFundMatter}
  J.~R.~Hiller, S.~Pinsky, Y.~Proestos, N.~Salwen and U.~Trittmann,
  Phys.~Rev.~D {\bf 76} (2007) 045008.
\bibitem{Kutasov94}
         {D.~Kutasov},
         {\em Nucl.~Phys.~}{\bf B414}
         (1994)
         33.
\end{thebibliography}
\end{document}